\documentclass[]{article}
\usepackage[a4paper, margin=1in]{geometry}
\usepackage{authblk}	
\usepackage{amsmath}    
\usepackage{graphicx}	
\usepackage{caption}	
\usepackage{subcaption}	
\usepackage[flushleft]{threeparttable}	
\usepackage{multirow}	

\usepackage[
	backend=bibtex,
	style=phys,
	sorting=none
]{biblatex}
\title{Forward-backward multiplicity distribution with the Chou-Yang model for $pp$ collisions at $\sqrt{s}=$ 0.9, 7 and 8 TeV from the CMS experiment}

\author{Z. Ong, P. Agarwal, H.W. Ang, A.H. Chan, C.H. Oh}
\affil{Department of Physics, National University of Singapore}
\date{}

\begin{document}

\maketitle

\begin{abstract}
	A Chou-Yang type multiplicity distribution comprising a total multiplicity component and a binomial asymmetry component is used to describe charged hadron multiplicity data at $\sqrt{s}=$ 0.9, 7 and 8 TeV from the CMS experiment at CERN. The data was obtained and processed from the CERN Open Data Portal. For the total multiplicity component, it was found that a convex sum of a Negative Binomial Distribution (NBD) and a Furry-Yule Distribution (FYD) is able to describe the shoulder-like structure characteristic of KNO scaling violation well. The mean cluster size produced from collisions was also found to increase with collision energy. A prediction is given for $pp$ collisions at $\sqrt{s}=$ 14 TeV.
\end{abstract}

\section{Introduction}

The Chou-Yang model was proposed by T.T. Chou and C.N. Yang in 1984~\cite{Chou:1984wp} to describe the forward-backward multiplicity distribution of high energy collisions. In their paper, they defined the variables
\begin{equation}
	n = n_\text{F} + n_\text{B},
\end{equation}
\begin{equation}
	z \equiv n_\text{F} - n_\text{B}
\end{equation}
to describe an event with total multiplicity $n$ and forward-backward (FB) asymmetry $z$ respectively. $n_\text{F}$ and $n_\text{B}$ are the number of particles produced in the forward ($\eta > 0$) and backward ($\eta < 0$) regions. They found that the $p\overline{p}$ collision data at $\sqrt{s}=$ 540 GeV from the UA5 collaboration seemed to obey
\begin{equation}
	\left[ \left< z^2 \right> \text{at fixed } n\right] = 2n,
	\label{eq:ChouYang2n}
\end{equation}
which led them to postulate that the asymmetry parameter $z$ is governed by a binomial distribution. Hence, they proposed that the $P(n,z)$ distribution be composed of two components -- one to describe the production of $n$ charged particles, and another to describe $z$ (i.e. how they are distributed in the FB direction):
\begin{equation}
	P(n,z) = \Psi \left( \frac{n}{\left< n \right>} \right) C_{n_F / 2}^{n/2} \left[ B \left( \frac{n}{2} \right) \right]^{-1},
\end{equation}
where $\Psi \left( \frac{n}{\left< n \right>} \right)$ is the KNO scaling function describing $n$, $C_{n_F / 2}^{n/2} = C_{(n+z)/4}^{n/2}$ is a combinatorial factor describing the binomial distribution of $z$, and $B$ is factor that normalises the latter,
\begin{equation}
	B \left( \frac{n}{2} \right) = \sum_{n_\text{F}} C_{n_{\text{F}}/2}^{n/2}
	\label{eq:ChouOriginalBinomialPart}
\end{equation}
such that the overall probability distribution $P(n,z)$ is normalised.

\subsection{Proposed modification to the Chou-Yang Model: total multiplicity $n$}

Presently, it is well-established that the multiplicity distributions of $pp$ collisions no longer obey KNO scaling~\cite{Alner:1985wj} at UA5 energies and beyond. This led to several studies that replaced the KNO scaling function in $P(n,z)$ with other analytic distributions. Lim \textit{et al.}~\cite{Lim:1989wt,Lim:1992ew} suggested replacing the KNO scaling function with the Negative Binomial Distribution (NBD, equation~\ref{eq:NBD}), and Lai \textit{et al.}~\cite{Lai:2009zz} and Phang \textit{et al.}~\cite{Phang:2019jut} developed the approach further by replacing the NBD with the Generalised Multiplicity Distribution (GMD, equation~\ref{eq:GMD}):

\begin{equation}	
	\label{eq:NBD}
	P_{\text{NBD}}(n;\overline{n},k)
	= \frac{\Gamma(n+k)}{\Gamma(n+1) \Gamma(k)}
	\left[ \frac{\overline{n}}{\overline{n}+k} \right]^{n}
	\left[ \frac{k}{\overline{n}+k} \right]^{k},
\end{equation}

\begin{equation}	
	\label{eq:GMD}
	P_{\text{GMD}}(n;\overline{n},k,k')
	= \frac{\Gamma(n+k)}{\Gamma(n-k'+1) \Gamma(k+k')}
	\left[ \frac{\overline{n}-k'}{\overline{n}+k} \right]^{n-k'}
	\left[ \frac{k+k'}{\overline{n}+k} \right]^{k+k'},
\end{equation}
where $\Gamma$ is the Gamma function, which generalises the arguments of the combinatorial prefactors of the NBD and GMD from positive integers to positive real numbers. Thus, we now have
\begin{equation}
	P(n,z) = P(n) C_{n_F / 2}^{n/2} \left[ B \left( \frac{n}{2} \right) \right]^{-1},
\end{equation}
where $P(n)$ is either $P_{\text{NBD}}(n)$ or $P_{\text{GMD}}(n)$.

The GMD is defined for values of $n$ such that $n \geq k'$. This mathematical requirement might severely constraint the overall ability of $P(n,z)$ to describe the data, since it would cause several bins in $P(n_\text{F}, n_\text{B})$ to be empty. Hence, we propose using a $P(n)$ constructed from a convex sum of the two limiting forms of the GMD (which becomes the NBD when $k'=0$, and the FYD when $k=0$), so that all values of $n$ can be described:
\begin{equation}
	P(n) = \alpha P_\text{NBD} (n; \overline{n}_\text{NBD}, k) + (1 - \alpha) P_\text{FYD}(n; \overline{n}_\text{FYD}, k'),
	\label{eq:NBD-FYD}
\end{equation}
where $\alpha \in \left[0,1\right]$.

\subsection{Mean cluster size $r$}

Lim \textit{et al.}~\cite{Lim:1989wt,Lim:1992ew} considered the prospect that the factor `2' in equation~\ref{eq:ChouOriginalBinomialPart} could take on other numbers. They replaced it with $r$, which results in a reinterpretation of the original binomial model: the original $n$ particles are first grouped into $r$ clusters, each of which then gets distributed in $z$ according to the modified Chou-Yang scheme. This has come to be known as the ``cluster model''; thus, in Chou and Yang's original paper~\cite{Chou:1984wp}, we have $r=2$ clusters for the special case of $p\overline{p}$ collisions at $\sqrt{s}=$ 540 GeV from UA5.

To this end, the overall Chou-Yang model has been modified to become
\begin{equation}
	P(n,z) = P(n) C_{n_F/r}^{n/r} \left[ B \left( \frac{n}{r} \right) \right]^{-1},
\end{equation}
where the mean cluster size $r$ is to be determined from experimental data. Equation~\ref{eq:ChouYang2n} also becomes modifed to be~\cite{Chou:1984wp,Lim:1989wt}
\begin{equation}
	\left[ \left< z^2 \right> \text{at fixed } n\right] = rn.
	\label{eq:ChouYang-rn}
\end{equation}

\section{Investigation at LHC energies}
We wish to make two measurements. The first is to investigate how our proposed NBD-FYD distribution (equation~\ref{eq:NBD-FYD}) is able to describe the collision data, and if it is able to offer some physical insight into the multiparticle production process.

The second measurement concerns the asymmetry part of the generalised version of the Chou-Yang model,
\begin{equation}
	P(z; r) = C_{n_F/r}^{n/r} \left[ B \left( \frac{n}{r} \right) \right]^{-1}.
\end{equation}
Specifically, we wish to find out how $r$ varies with collision energy in the CMS Run 1 data. This will be obtained by extracting the 2-dimensional probability distribution $P(n_\text{F}, n_\text{B})$ from data, plotting $\left< z^2 \right>$ vs. $n$, and performing linear regression to obtain $r$ in equation~\ref{eq:ChouYang-rn}.

\section{About the data}

This analysis is performed on Run 1 data from the CMS collaboration processed from the CMS Open Data Portal, covering centre-of-mass energies $\sqrt{s}=$ 0.9, 7 and 8 TeV. The analysis method follows largely that of CMS~\cite{Khachatryan:2010nk}, which analysed minimum-bias (MinBias), non-single diffractive (NSD) multiplicity distributions.

NSD events were selected by requiring that at least one forward hadron (HF) calorimeter tower on
each side of the detector have at least 3 GeV of energy deposited in the event. The primary vertex was chosen as the vertex with the highest number of associated tracks, which must also be within 15 cm of the reconstructed beamspot in the beam axis and be of good reconstruction quality (ndof $>$ 4).

Good quality tracks were selected by requiring them to carry the \texttt{highPurity} label. Furthermore, we select for tracks with $<$10\% relative error on the transverse momentum ($p_{\text{T}}$) measurement ($\sigma_{p_{\text{T}}} / p_{\text{T}} < 0.1$) to reject low-quality and badly reconstructed tracks. Secondaries were removed by requiring a small impact parameter with respect to the selected primary vertex. Also, tracks were required to have $p_{\text{T}} > 500$ MeV/c, which will be extrapolated to zero via unfolding.

Finally, unfolding was performed using an iterative ``Bayesian unfolding method'', which is more accurately known as ``D’Agostini iteration with early stopping'' and described in~\cite{DAgostini:1994fjx}. This infers the original charged hadron multiplicity distribution (MinBias NSD) from the charged track multiplicity distribution measured.

Tables~\ref{tab:RECOdatasets} and~\ref{tab:MCdatasets} in the Appendix summarise the datasets used.

\section{Results and discussion}

\subsection{Multiplicity component}

\begin{table}
	\centering
	\begin{threeparttable}
		\caption{Best-fit parameters for NBD-FYD model}
		\label{tab:NBD-FYD parameters}
		\begin{tabular}{c|ccc}
			\hline
			$\sqrt{s}$ & 900 GeV & 7 TeV & 8 TeV \\
			\hline
			$\alpha$ & $0.541\pm0.001$ & $0.814\pm0.001$ & $0.846\pm0.002$ \\
			$\overline{n}_\text{NBD}$ & $32.18\pm0.04$ & $29.56\pm0.04$ & $43.2\pm0.1$\\
			$k$ & $5.60\pm0.02$ & $1.108\pm0.003$ & $0.886\pm0.003$\\
			$\overline{n}_\text{FYD}$ & $15.09\pm0.02$ & $15.15\pm0.04$ & $14.16\pm0.06$\\
			$k'$ & $4.9999904\pm0.0000008$ & $3.99963\pm0.00002$ & $2.81\pm0.03$\\
			\hline
			$\chi^{2}$/d.o.f. & 60.55 & 11.44 & 13.76 \\
			\hline
		\end{tabular}
		\begin{tablenotes}
			\small
			\item Note: The large number of decimal figures for $k'$ for $\sqrt{s}=$ 900 GeV and 7 TeV are the optimiser's attempts to generate as large a real number below 5 and 4 respectively, so as not to sacrifice one data point from the FYD component (which is defined only for $n \geq k'$).
		\end{tablenotes}
	\end{threeparttable}
\end{table}

\begin{figure}
	\centering
	\begin{subfigure}[b]{0.6\textwidth}
		\centering
		\includegraphics[width=\textwidth]{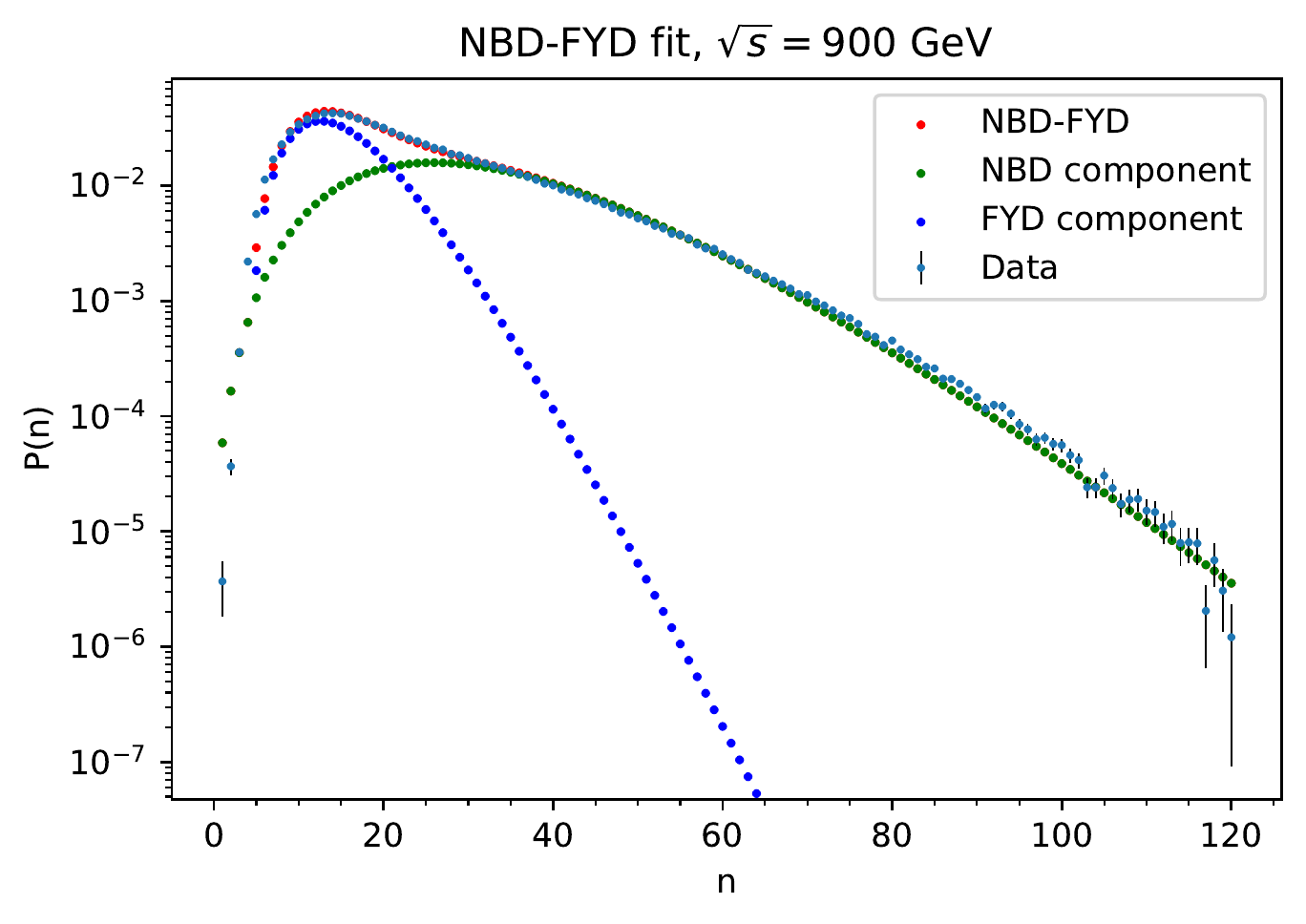}
		\caption{900 GeV}
		\label{fig:NBD-FYD_900GeV}
	\end{subfigure}
	\par\bigskip
	\begin{subfigure}[b]{0.6\textwidth}
		\centering
		\includegraphics[width=\textwidth]{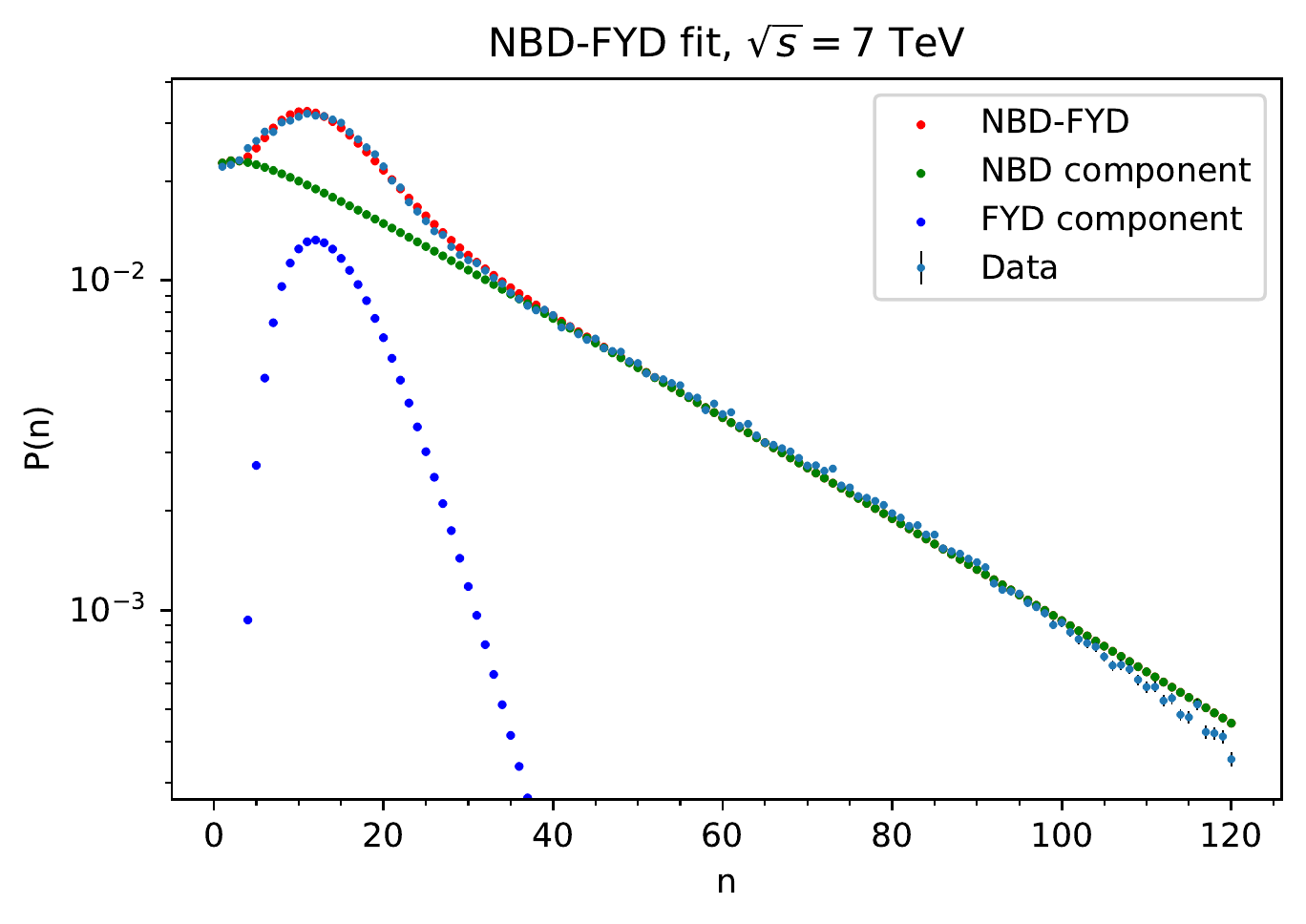}
		\caption{7 TeV}
		\label{fig:NBD-FYD_7TeV}
	\end{subfigure}
	\par\bigskip
	\begin{subfigure}[b]{0.6\textwidth}
		\centering
		\includegraphics[width=\textwidth]{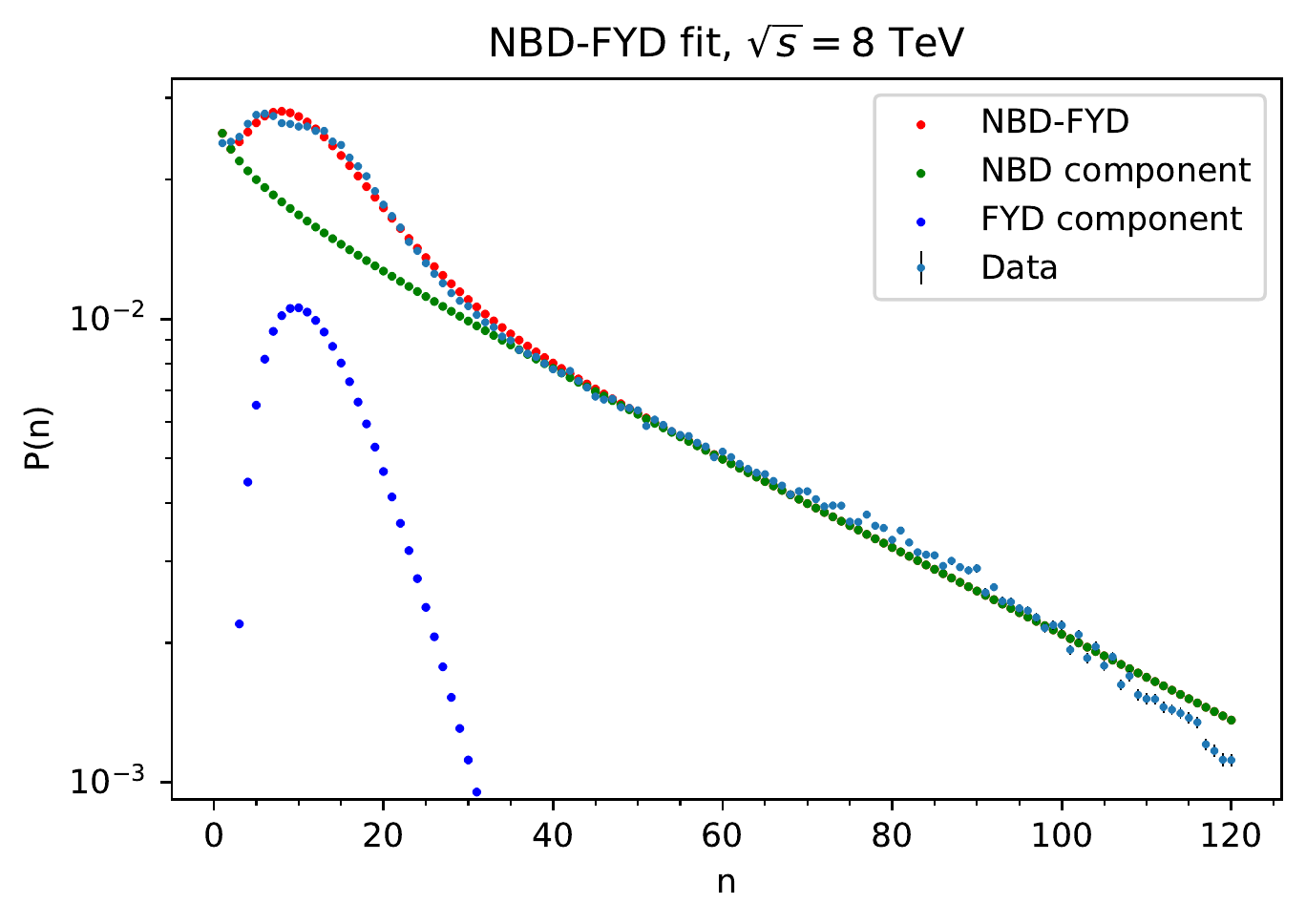}
		\caption{8 TeV}
		\label{fig:NBD-FYD_8GeV}
	\end{subfigure}
	\caption{Fits of the NBD-FYD model to data for $1 \leq n \leq 120$, with their NBD and FYD components also shown.}
	\label{fig:NBD-FYD fits}
\end{figure}

Table~\ref{tab:NBD-FYD parameters} summarises the best-fit parameters found for the NBD-FYD model in describing the data, and Figure~\ref{fig:NBD-FYD fits} shows the collision data, the fitted NBD-FYD model and its constituent NBD and FYD components. In optimising the parameters, only data points within $1 \leq n \leq 120$ were used; $n=0$ was excluded from the fitting due to the high values of $P(0)$ which are not described by the model, while data points at $n > 120$ suffered from low statistics and were not smooth, which would interfere adversely with the optimisation process. The $\chi^2/\text{d.o.f.}$ values are much greater than unity due to the small magnitudes of the errors (due to underestimating the systematic errors). However, it can be ascertained visually that the model describes the data very well. Most importantly, the model is able to describe the shoulder-like structure (from KNO scaling violation) of the data.

It is interesting to note the manner in which the NBD-FYD model brings about the shoulder-like structure. The NBD component (green dots in Figure~\ref{fig:NBD-FYD fits}) provides the general overall downward-sloping profile, and the FYD component (blue dots) provides the protruding ``head'', which also defines the ``shoulder''. As $\sqrt{s}$ increases, the FYD component becomes narrower to manifest a more pronounced change in gradient (i.e. a more pronounced shoulder-like structure). This narrowing is brought about by an increase in $\alpha$ and a decrease in $k'$. It is also worth noting that $\overline{n}_\text{FYD}$ remains relatively constant at all values of $\sqrt{s}$.

\subsubsection*{Interpretation of NBD-FYD model}

The above findings raise an issue of interpretation that is potentially incompatible with the GMD. Chan and Chew's analysis~\cite{Chan:1990hs} showed that as collision energies increase towards the TeV scale, the GMD will approach the FYD (via decreasing $k$ and increasing $k'$), indicating significant contribution from gluon branching. The NBD-FYD model proposed here seems to suggest the opposite, with a diminishing FYD component. While the NBD-FYD model describes the data excellently, it seems to be incompatible with the GMD's notion of quark and gluon branching from the NBD and FYD components respectively.

It is useful to reconsider the origins of the NBD and FYD. As Hwa pointed out~\cite{Hwa:1987kc}, these are particular solutions to the basic evolution equation
\begin{equation}
	\frac{\partial P_n (t)}{\partial t}  = a_{n+1} P_{n+1} + c_{n-1} P_{n-1} - (a_n + c_n) P_n,
\end{equation}
which has been adapted to describe the evolution of the multiplicity distribution $P_n(t)$, with $t$ as the QCD evolution parameter. The coefficient $c_n$ is proportional to ``creation'' processes, which describes birth, emission, fragmentation or branching in the context of multiparticle production. $a_n$ is proportional to ``annihilation'' processes, which describes death, absorption, recombination or decay~\cite{Hwa:1987kc}.

In this framework, the NBD is the solution to the evolution equation describing birth, death and immigration processes, having coefficients of the form $a_n = \alpha n$, $c_n = \beta n + \gamma$. The FYD is the solution describing only birth processes, having coefficients of the form $a_n = 0$, $c_n = \beta n$.

When considered with the results in Table~\ref{tab:NBD-FYD parameters}, the NBD-FYD model paints a physical picture of an increasingly complex mode of multiparticle production. At lower energies, multiparticle production mainly involves pure-birth processes, as described by the FYD; at higher energies, additional death and immigration processes come into play, as described by the NBD.

\subsection{FB component}

\begin{figure}
	\centering
	\begin{subfigure}[b]{0.49\textwidth}
		\centering
		\includegraphics[width=\textwidth]{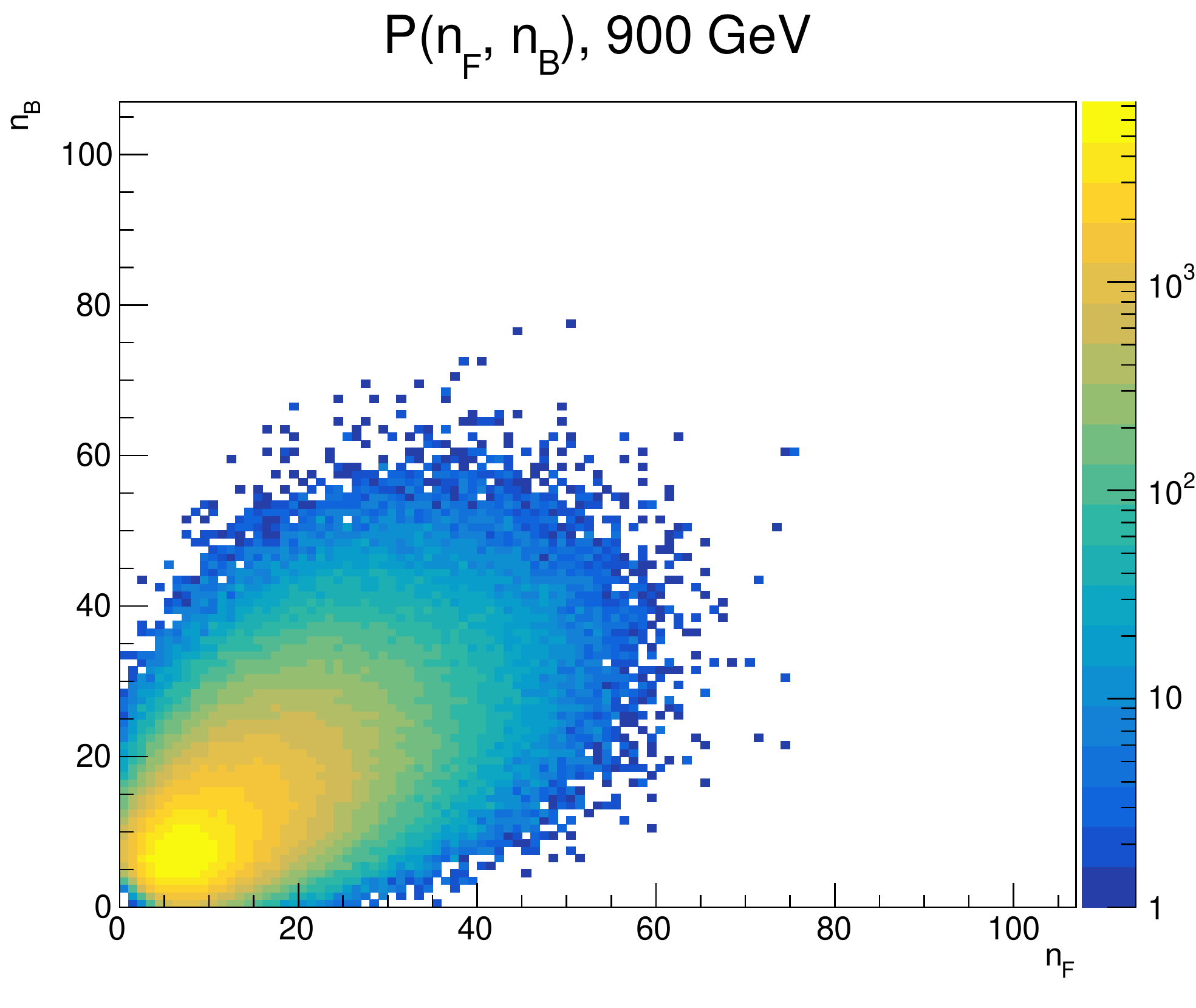}
		\caption{900 GeV}
		\label{fig:P(nF,nB)_900GeV}
	\end{subfigure}
	\hfill
	\begin{subfigure}[b]{0.49\textwidth}
		\centering
		\includegraphics[width=\textwidth]{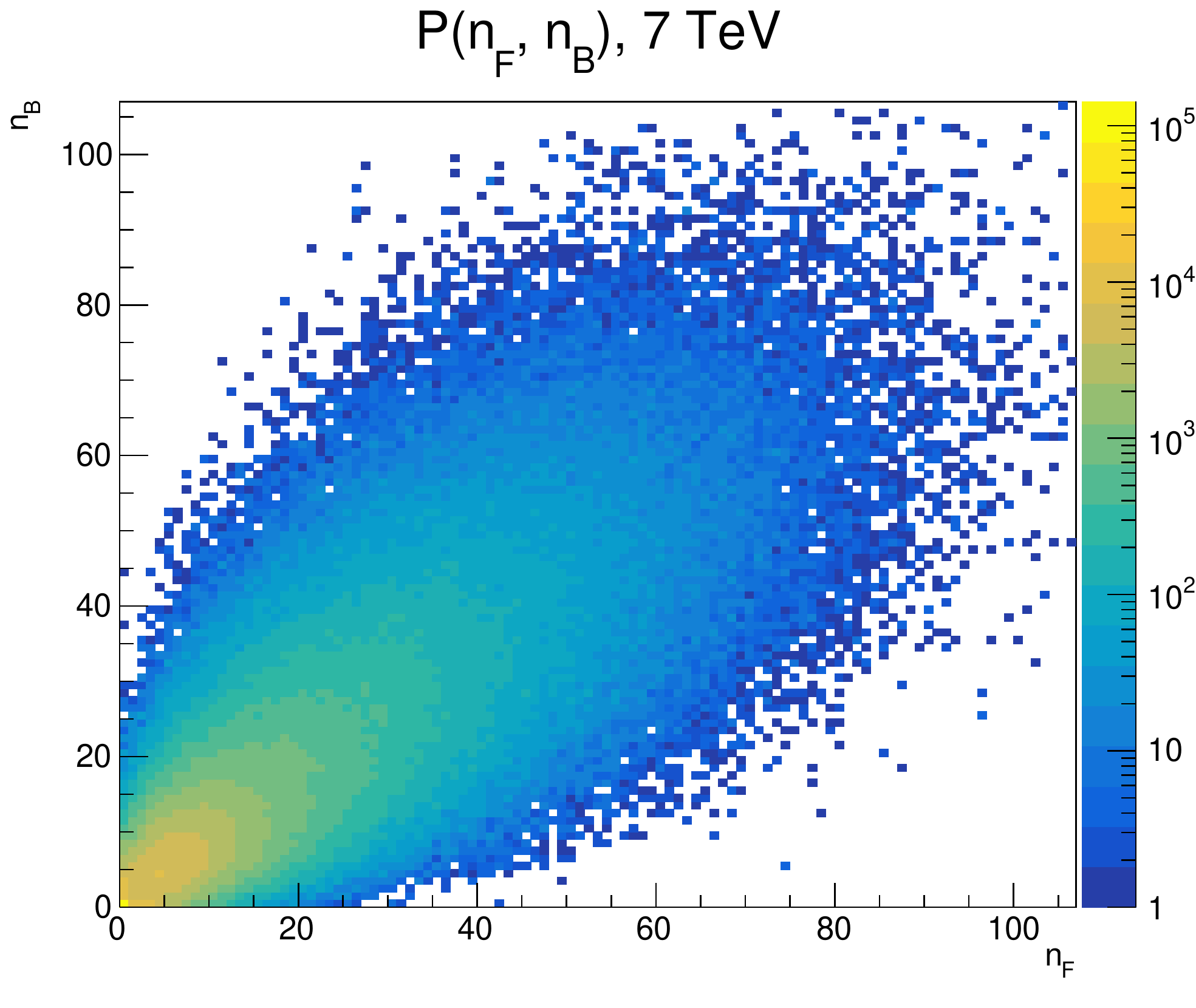}
		\caption{7 TeV}
		\label{fig:P(nF,nB)_7TeV}
	\end{subfigure}
	\par\bigskip
	\begin{subfigure}[b]{0.49\textwidth}
		\centering
		\includegraphics[width=\textwidth]{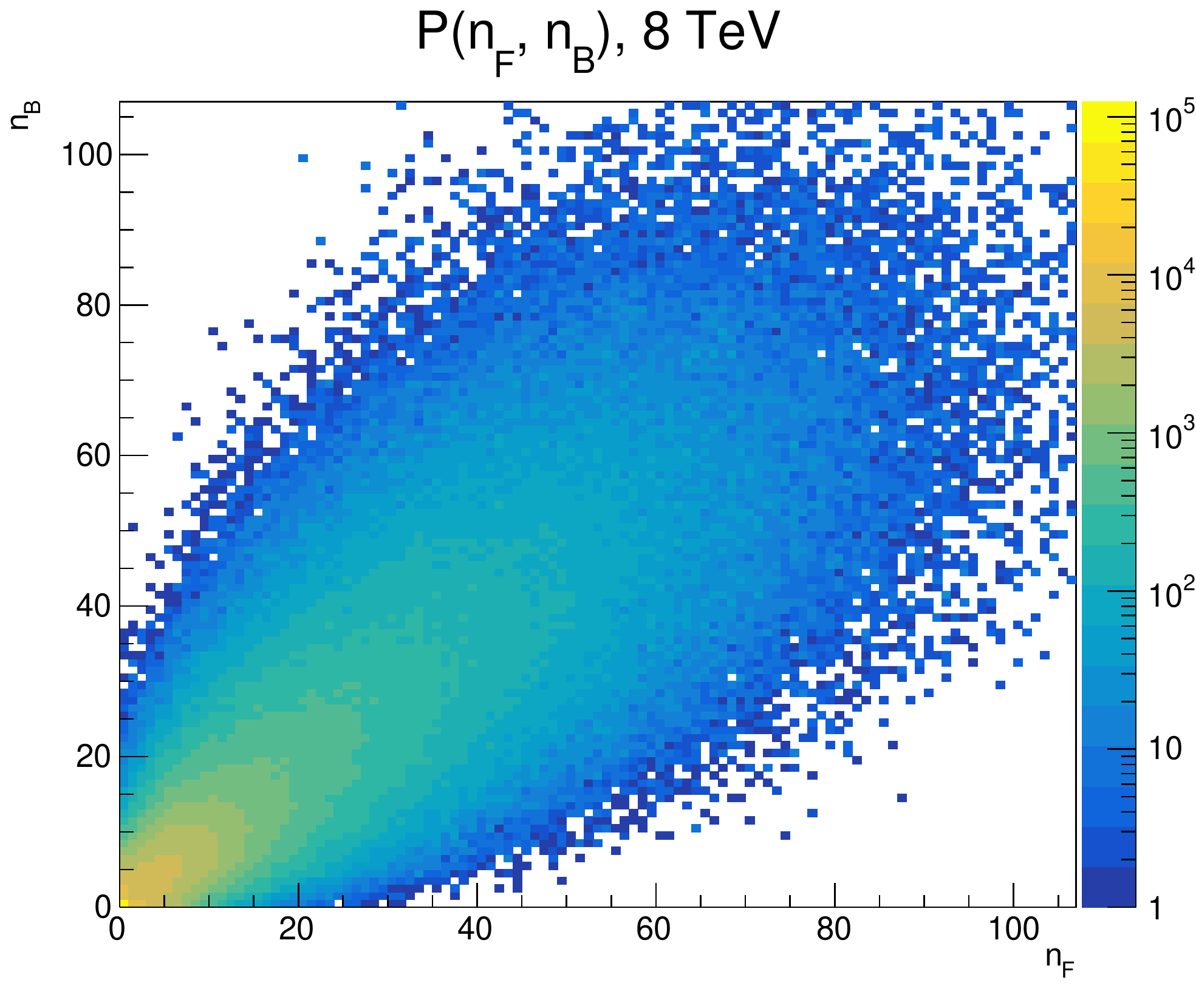}
		\caption{8 TeV}
		\label{fig:P(nF,nB)_8GeV}
	\end{subfigure}
	\caption{2D multiplicity distribution $P(n_\text{F}, n_\text{B})$ for $\sqrt{s}=$ 900 GeV, 7 TeV and 8 TeV, with the colour legend in log scale.}
	\label{fig:P(nF,nB)}
\end{figure}

\begin{figure}
	\centering
	\begin{subfigure}[b]{0.45\textwidth}
		\centering
		\includegraphics[width=\textwidth]{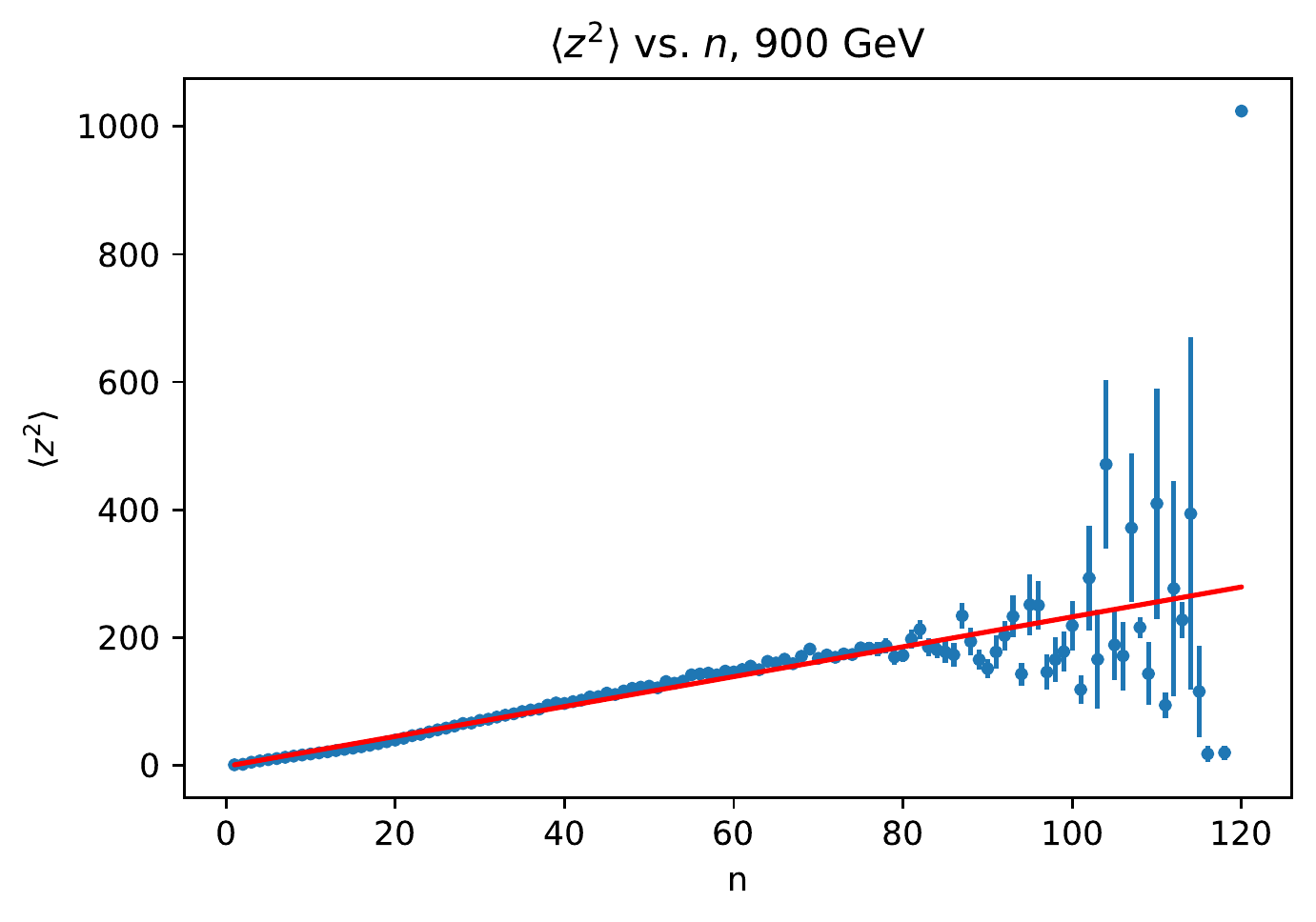}
		\caption{900 GeV}
		\label{fig:FBcorrelation_<z^2>vsN_900GeV}
	\end{subfigure}
	\hfill
	\begin{subfigure}[b]{0.45\textwidth}
		\centering
		\includegraphics[width=\textwidth]{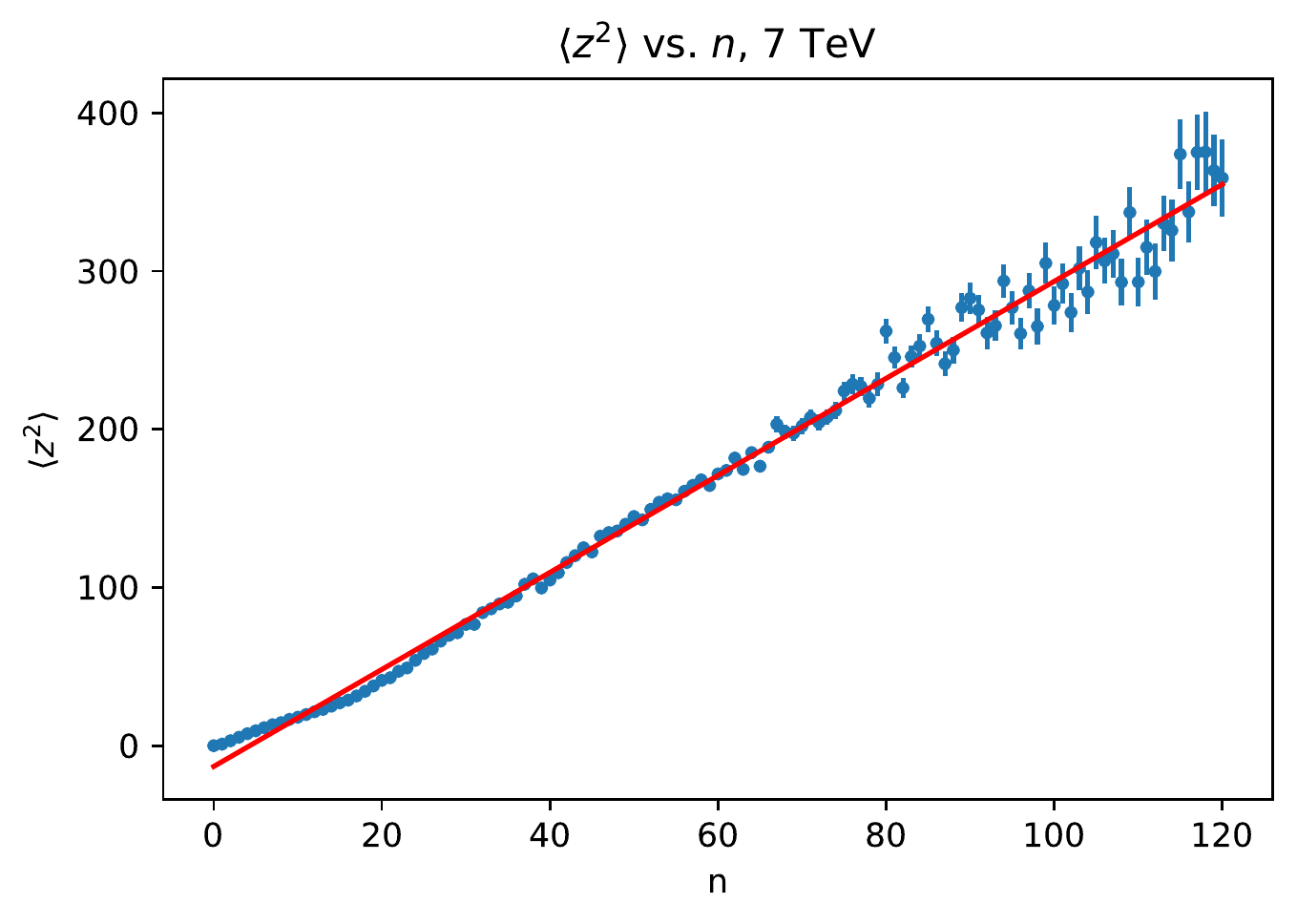}
		\caption{7 TeV}
		\label{fig:FBcorrelation_<z^2>vsN_7TeV}
	\end{subfigure}
	\par\bigskip
	\begin{subfigure}[b]{0.45\textwidth}
		\centering
		\includegraphics[width=\textwidth]{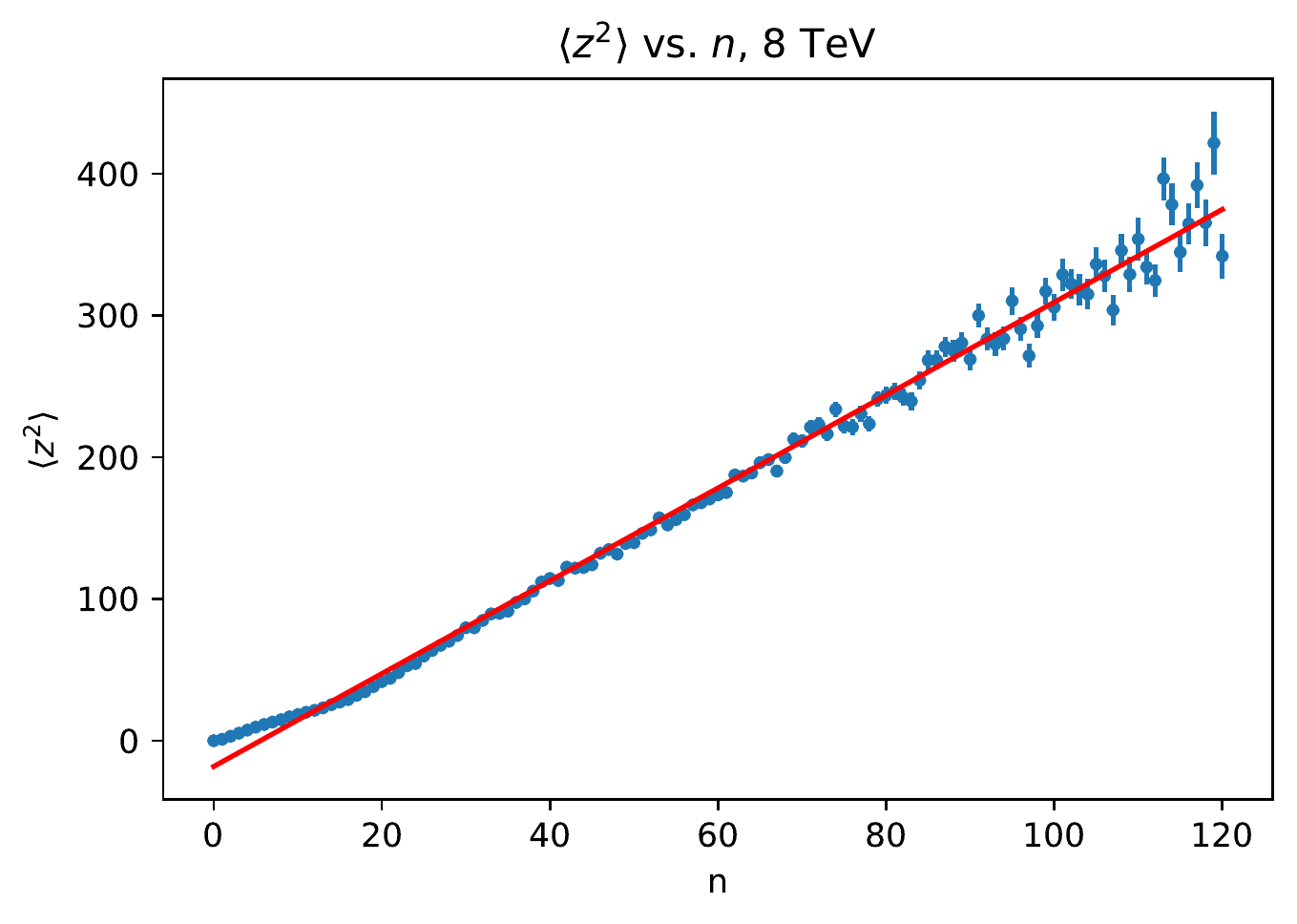}
		\caption{8 TeV}
		\label{fig:FBcorrelation_<z^2>vsN_8GeV}
	\end{subfigure}
	\caption{$\left< z^2 \right>$ vs. $n$ for $\sqrt{s}=$ 900 GeV, 7 TeV and 8 TeV. The slope of the best-fit line is equal to the mean cluster size $r$. Data points have been included for $n \leq 120$ to avoid truncation effects from the finiteness of $P(n_\text{F}, n_\text{B})$.}
	\label{fig:FBcorrelation_<z^2>vsN}
\end{figure}

\begin{figure}
	\centering
	\includegraphics[width=.6\textwidth]{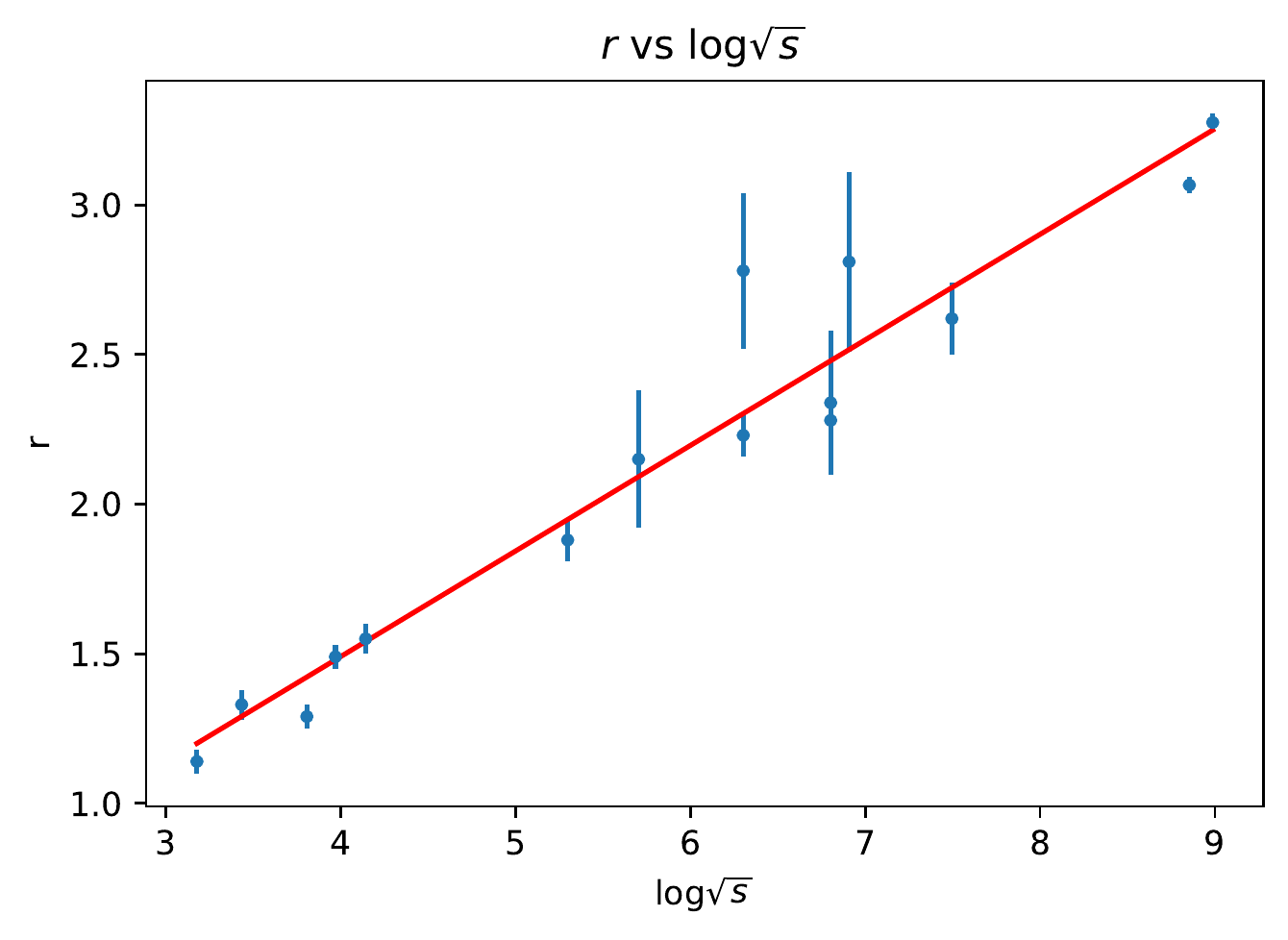}
	\caption{Variation of mean cluster size $r$ as a function of $\log \sqrt{s}$, for hadronic collisions with $\sqrt{s}$ ranging from 24 GeV to 8 TeV.}
	\label{fig:FB_r_vs_energy}
\end{figure}

Figure~\ref{fig:P(nF,nB)} shows the unfolded (and unnormalised) 2D multiplicity distribution plots of the data\footnote{The maximum values of $n_\text{F}$ and $n_\text{B}$ were restricted to 107 due to the memory limitations (in computer RAM) of the ROOT system. This will affect the computation of high values of total multiplicity $n$ due to truncation effects.}. From this, $\left< z^2 \right>$ vs. $n$ can be computed, and the results are shown in Figure~\ref{fig:FBcorrelation_<z^2>vsN}. The values of $r$ obtained from the data are summarised in Table~\ref{tab:FBclusterSize_r_results} and compiled alongside results from previous studies. Figure~\ref{fig:FB_r_vs_energy} shows how $r$ varies with collision energy $\sqrt{s}$ for hadronic collisions. The increase in $r$ seems to follow the relation~\cite{Lai:2009zz}
\begin{equation}
	r = \alpha \log \sqrt{s} + \beta,
\end{equation}
with $\alpha = 0.34 \pm 0.03$ and $\beta = 0.2 \pm 0.2$.

\begin{table}
	\centering
	\caption{Measured mean cluster size $r$ for hadronic collisions, compiled with previous studies at other energies}
	\begin{tabular}{ cccc }
		\hline
		Collaboration & $\sqrt{s}\text{ (GeV)}$ & $r$ & Source\\
		\hline
		\multirow{5}{*}{ISR ($pp$)}
		& 24		& $1.14 \pm 0.04$
		& \multirow{5}{*}{\cite{Lim:1989wt}}\\
		& 31		& $1.33 \pm 0.05$ \\
		& 45		& $1.29 \pm 0.04$ \\
		& 53		& $1.49 \pm 0.04$ \\
		& 63		& $1.55 \pm 0.05$ \\
		\hline
		\multirow{3}{*}{UA5 ($p\overline{p}$)}
		& 200		& $1.88 \pm 0.07$
		& \multirow{3}{*}{\cite{Lim:1989wt}}\\
		& 546		& $2.23 \pm 0.07$ \\
		& 900		& $2.28 \pm 0.07$ \\
		\hline
		\multirow{4}{*}{E735 ($p\overline{p}$)}
		& 300		& $2.15 \pm 0.23$
		& \multirow{4}{*}{\cite{E735:1995ktc}}\\
		& 546		& $2.78 \pm 0.26$ \\
		& 1000	& $2.81 \pm 0.30$ \\
		& 1800	& $2.62 \pm 0.12$ \\
		\hline
		\multirow{3}{*}{CMS ($pp$)}
		& 900		& $2.3 \pm 0.2$
		& \multirow{3}{*}{(New measurement)}\\
		& 7000	& $3.07 \pm 0.03$ \\
		& 8000	& $3.28 \pm 0.03$ \\
		\hline
	\end{tabular}	
	\label{tab:FBclusterSize_r_results}
\end{table}

\section{Conclusion}

We presented a model that describes the multiplicity distribution as 
a convex sum of a Furry-Yule Distribution (FYD) and a Negative Binomial Distribution (NBD). It was found to describe data well. The FYD component diminishes with increasing energy while the NBD component becomes more prominent, the latter suggesting an increase in complexity of multiparticle production with death and immigration processes starting to emerge.

An increase in mean cluster size $r$ has also been observed in hadronic collisions from $\sqrt{s}=$ 24 GeV to 8 TeV. The increase in $r$ seems to follow the relation $r = \alpha \log \sqrt{s} + \beta$, with $\alpha = 0.35 \pm 0.03$ and $\beta = 0.1 \pm 0.2$. Based on this, we predict that we would have $r = 3.4 \pm 0.5$ at $\sqrt{s}=$ 14 TeV, and we look forward to analysing 14 TeV data when it becomes available.

\section*{Acknowledgements}
The authors would like to thank Prof Kati Lassila-Perini from the Data Preservation and Open Access project in the CMS experiment at CERN for help with the 8 TeV Monte Carlo dataset. This work is supported by the NUS Research Scholarship.

\newpage
\appendix

\section{Datasets used}

\begin{table}[h!]
	\centering
	\caption{Summary of CMS collider datasets used from CERN Open Data Portal}
	\begin{tabular}{ |c|l|c| } 
		\hline
		$\sqrt{s}$  & \multirow{2}{*}{Dataset} & \multirow{2}{*}{Ref.} \\
		(TeV) & & \\
		\hline \hline
		0.9	& /MinimumBias/Commissioning10-07JunReReco\_900GeV/RECO & \cite{data900GeV} \\
		\hline
		7	& /MinimumBias/Run2010A-Apr21ReReco-v1/AOD & \cite{data7TeV} \\ 
		\hline
		8	& /MinimumBias/Run2012B-22Jan2013-v1/AOD & \cite{data8TeV} \\ 
		\hline	
	\end{tabular}	
	\label{tab:RECOdatasets}
\end{table}

\begin{table}[h!]
	\centering
	\caption{Summary of Monte Carlo datasets used from CERN Open Data Portal}
	\begin{tabular}{ |c|l|c| } 
		\hline
		$\sqrt{s}$  & \multirow{2}{*}{Dataset} & \multirow{2}{*}{Ref.} \\
		(TeV) & & \\
		\hline \hline
		
		\multirow{2}{*}{0.9}	& /MinBias\_TuneZ2\_900GeV\_pythia6\_cff\_py & \multirow{2}{*}{\cite{mc900GeV}}\\
		& \_GEN\_SIM\_START311\_V2\_Dec11\_v2 &  \\
		\hline
		
		\multirow{2}{*}{7}	& /MinBias\_TuneZ2star\_7TeV\_pythia6/Summer12-LowPU2010 & \multirow{2}{*}{\cite{mc7TeV}} \\ 
		& \_DR42-PU\_S0\_START42\_V17B-v1/AODSIM &  \\ 
		\hline
		
		\multirow{2}{*}{8}	& /MinBias\_TuneZ2star\_8TeV-pythia6/Summer12\_DR53X-PU & \multirow{2}{*}{\cite{mc8TeV}} \\
		& \_S10\_START53\_V7A-v1/AODSIM & \\
		\hline	
	\end{tabular}	
	\label{tab:MCdatasets}
\end{table}

\printbibliography

@article{data900GeV,
	author = "{CMS collaboration (2019)}",
	title = "{MinimumBias primary dataset in RECO format from the 0.9 TeV Commissioning run of 2010 (/MinimumBias/Commissioning10-07JunReReco\_900GeV/RECO). CERN Open Data Portal.}",
	doi = "DOI:10.7483/OPENDATA.CMS.1R58.OMBD"
}

@article{data7TeV,
	author = "{CMS collaboration (2019)}",
	title = "{MinimumBias primary dataset in AOD format from RunA of 2010\\ (/MinimumBias/Run2010A-Apr21ReReco-v1/AOD). CERN Open Data Portal.}",
	doi = "DOI:10.7483/OPENDATA.CMS.6B3H.TR6Z"
}

@article{data8TeV,
	author = "{CMS collaboration (2017)}",
	title = "{MinimumBias primary dataset in AOD format from RunB of 2012\\ (/MinimumBias/Run2012B-22Jan2013-v1/AOD). CERN Open Data Portal.}",
	doi = "DOI:10.7483/OPENDATA.CMS.HU6U.DRLD"
}

@article{mc900GeV,
	author = "{CMS Collaboration (2019)}",
	title = "{Simulated dataset\\ MinBias\_TuneZ2\_900GeV\_pythia6\_cff\_py\_GEN\_SIM\_START311\_V2\_Dec11\_v2 in GEN-SIM-RECO format for 2010 commissioning data. CERN Open Data Portal.}",
	doi = "DOI:10.7483/OPENDATA.CMS.JPB5.X7CN"
}

@article{mc7TeV,
	author = "{CMS Collaboration (2018)}",
	title = "{Simulated dataset MinBias\_TuneZ2star\_7TeV\_pythia6 in AODSIM format for 2010 collision data. CERN Open Data Portal.}",
	doi = "DOI:10.7483/OPENDATA.CMS.VTJ2.E5JN"
}

@article{mc8TeV,
	author = "{CMS Collaboration (2021)}",
	title = "{Simulated dataset MinBias\_TuneZ2star\_8TeV-pythia6 in AODSIM format for 2012 collision data. CERN Open Data Portal.}",
	doi = "DOI:10.7483/OPENDATA.3GIM.7SPW"
}

@article{Alner:1985wj,
	author = "Alner, G. J. and others",
	collaboration = "UA5",
	title = "{Scaling Violations in Multiplicity Distributions at 200 GeV and 900 GeV}",
	reportNumber = "CERN-EP/85-197",
	doi = "10.1016/0370-2693(86)91304-3",
	journal = "Phys. Lett. B",
	volume = "167",
	pages = "476--480",
	year = "1986"
}

@article{Chan:1990hs,
	author = "Chan, A. H. and Chew, C. K.",
	title = "{Parton branching model for $p\overline{p}$ collisions}",
	doi = "10.1103/PhysRevD.41.851",
	journal = "Phys. Rev. D",
	volume = "41",
	pages = "851--862",
	year = "1990"
}

@article{Chou:1984wp,
	author = "Chou, T. T. and Yang, Chen Ning",
	title = "{Binomial Distribution for the Charge Asymmetry Parameter}",
	reportNumber = "Print-84-0079 (GEORGIA)",
	doi = "10.1016/0370-2693(84)90478-7",
	journal = "Phys. Lett. B",
	volume = "135",
	pages = "175--178",
	year = "1984"
}

@article{DAgostini:1994fjx,
	author = "D'Agostini, G.",
	title = "{A Multidimensional unfolding method based on Bayes' theorem}",
	reportNumber = "DESY-94-099",
	doi = "10.1016/0168-9002(95)00274-X",
	journal = "Nucl. Instrum. Meth. A",
	volume = "362",
	pages = "487--498",
	year = "1995"
}

@article{E735:1995ktc,
	author = "Alexopoulos, T. and others",
	collaboration = "E735",
	title = "{Charged particle multiplicity correlations in $p\overline{p}$ collisions at $\sqrt{s}=$ 0.3-1.8 TeV}",
	reportNumber = "FERMILAB-PUB-95-136-E",
	doi = "10.1016/0370-2693(95)00554-X",
	journal = "Phys. Lett. B",
	volume = "353",
	pages = "155--160",
	year = "1995"
}

@article{Hwa:1987kc,
	author = "Hwa, Rudolph C.",
	title = "{Branching Processes in Multiparticle Production}",
	reportNumber = "OITS-374",
	doi = "10.1142/9789814503259_0014",
	journal = "Adv. Ser. Direct. High Energy Phys.",
	volume = "2",
	pages = "556--613",
	year = "1988"
}

@article{Khachatryan:2010nk,
	author = "Khachatryan, Vardan and others",
	collaboration = "CMS",
	title = "{Charged Particle Multiplicities in $pp$ Interactions at $\sqrt{s}=0.9$, 2.36, and 7 TeV}",
	eprint = "1011.5531",
	archivePrefix = "arXiv",
	primaryClass = "hep-ex",
	reportNumber = "CERN-PH-EP-2010-048, CMS-QCD-10-004",
	doi = "10.1007/JHEP01(2011)079",
	journal = "JHEP",
	volume = "01",
	pages = "079",
	year = "2011"
}

@article{Lai:2009zz,
	author = "Lai, W. C. and Chan, A. H. and Oh, C. H.",
	title = "{Chou-Yang multiplicity correlations in high energy multiparticle production and LHC prediction}",
	doi = "10.1142/S0217751X09047181",
	journal = "Int. J. Mod. Phys. A",
	volume = "24",
	pages = "3552--3560",
	year = "2009"
}

@article{Lim:1989wt,
	author = "Lim, S. L. and Lim, Y. K. and Oh, C. H. and Phua, K. K.",
	title = "{Forward-backward multiplicity correlation in high energy hadron-hadron collisions}",
	reportNumber = "Print-89-0256 (SINGAPORE), NUS-HEP-82-02",
	doi = "10.1007/BF01550941",
	journal = "Z. Phys. C",
	volume = "43",
	pages = "621",
	year = "1989"
}

@article{Lim:1992ew,
	author = "Lim, S. L. and Oh, C. H. and Phua, K. K.",
	title = "{The forward-backward multiplicity correlations, the single hemisphere multiplicity distributions and the mean cluster size in high-energy $e^{+}e^{-}$ and hadron-hadron collisions}",
	reportNumber = "NUS-HEP-08-91",
	doi = "10.1007/BF01881713",
	journal = "Z. Phys. C",
	volume = "54",
	pages = "107--114",
	year = "1992"
}

@article{Phang:2019jut,
	author = "Phang, S. W. and Chan, A. H. and Oh, C. H. and Yuen, E. and Ong, Z. and Leong, Q. X.",
	title = "{Chou-Yang Model for Forward-Backward Multiplicity Correlations at 7 TeV using Generalized Multiplicity Distribution.}",
	doi = "10.1051/epjconf/201920609005",
	journal = "EPJ Web Conf.",
	volume = "206",
	pages = "09005",
	year = "2019"
}

\end{document}